\pdfoutput=1
\documentclass[aip,rsi,
reprint,
graphicx,amsmath,amssymb
]{revtex4-1} 

\usepackage[utf8]{inputenc}  

\usepackage{subcaption}
\usepackage{caption}
\usepackage{letltxmacro}

\usepackage[T1]{fontenc}    
\usepackage{lmodern}        

\usepackage[table,xcdraw]{xcolor}  
\usepackage{soulutf8}  
\usepackage{ulem}  

\usepackage{graphicx}
\usepackage{dcolumn}
\usepackage{bm}

\usepackage{epsfig}
\usepackage{hyperref}
\usepackage{natbib}
\usepackage{float}

\definecolor{blue}{RGB}{      31, 119, 180 }
\definecolor{orange}{RGB} {  255, 127,  14 }
\definecolor{red}{RGB}{      255,   0,   0 }
\definecolor{darkred}{RGB}{  128,   0,   0 }
\definecolor{turquoise}{RGB}{  0, 128, 128 }
\definecolor{green}{RGB}{     44, 160,  44 }
\definecolor{gray}{RGB}{     128, 128, 128 }

\newcommand{\blue}[0]{   \textcolor{blue}{blue} }
\newcommand{\orange}[0]{ \textcolor{orange}{orange} }
\newcommand{\red}[0]{    \textcolor{red}{red} }
\newcommand{\green}[0]{  \textcolor{green}{green} }
\newcommand{\gray}[0]{   \textcolor{gray}{gray} }

\begin{document}
\title{Compact embedded device for lock-in measurements and experiment active control}

\author{Marcelo Alejandro \surname{Luda}}
\email[]{mluda@citedef.gob.ar}
\affiliation{CEILAP, CITEDEF, J.B. de La Salle 4397, 1603 Villa Martelli, Buenos Aires, Argentina}

\author{Martin \surname{Drechsler}}
\affiliation{Departamento de F\'{\i}sica \& IFIBA, FCEyN, UBA, Pabell\'on 1, Ciudad Universitaria, 1428 Buenos Aires, Argentina}

\author{Christian Tom\'as \surname{Schmiegelow}}
\affiliation{Departamento de F\'{\i}sica \& IFIBA, FCEyN, UBA, Pabell\'on 1, Ciudad Universitaria, 1428 Buenos Aires, Argentina}

\author{Jorge \surname{Codnia}}
\affiliation{CEILAP, CITEDEF, J.B. de La Salle 4397, 1603 Villa Martelli, Buenos Aires, Argentina}

\begin{abstract}
We present a multi-purpose toolkit for digital processing, acquisition and feedback control designed for
physics labs. The kit provides in a compact device the functionalities of
several instruments: function generator, oscilloscope, lock-in amplifier,
proportional-integral-derivative filters, Ramp scan generator and a Lock-control.
The design combines Field-Programmable-Gate-Array processing and microprocessor programing to
get precision, ease of use and versatility. It can be remotely operated through the network with
different levels of control: from simple out-of-the-shelve Web GUI to remote script control or
in-device programmed operation. Three example applications are presented in this work on laser
spectroscopy and laser locking experiments. The examples includes side-fringe locking, peak locking
through lock-in demodulation, complete in-device Pound–Drever–Hall modulation and demodulation
at 31.25~MHz and advanced acquisition examples like real-time data streaming for remote storage.
\end{abstract}



\maketitle

\section{Introduction} 

The demand for stabilization systems in optics, atomic and molecular physics experiments is increasing
more and more with time. The realization of high accuracy measurements relies often
on the fine control of several variables at a time in the same experiment: light intensities, emission
wavelength, temperature, modulation depth, mechanical stress or position of elements; are just some examples
of variables that need to be controlled for that purpose.

The standard approach in control theory for a stabilization and control procedure is a
feedback scheme\cite{RMP_Bechhoefer_2005}.
This is accomplished by measuring certain variables of the system-under-control and
acting on it with a set of control-parameters.
The most common procedure is to measure outcomes, compare them to some preset desired values
and then act on the system with a chosen strategy to compensate the deviations.
The realization of this strategy is the core of the control system and its instrumentation has evolved from welded
electronics analog circuits to more sophisticated and versatile digital processing systems. Probably,
the most popular strategy used for control is the proportional-integral-derivative (PID) filter.
Though it is known not to be the optimal strategy for all systems, its ease of use and understanding,
as well as its satisfactory results makes it the usual workhorse in most practical setups and the one here chosen.
Moreover, one can combine various PID filters to control multiple inputs and multiple outputs (MIMO), add filters
and signal processing to any stage making the uses of PID control
a versatile and intuitive platform for control.

\begin{figure*}[ht]
  \begin{center}
    \epsfig{file=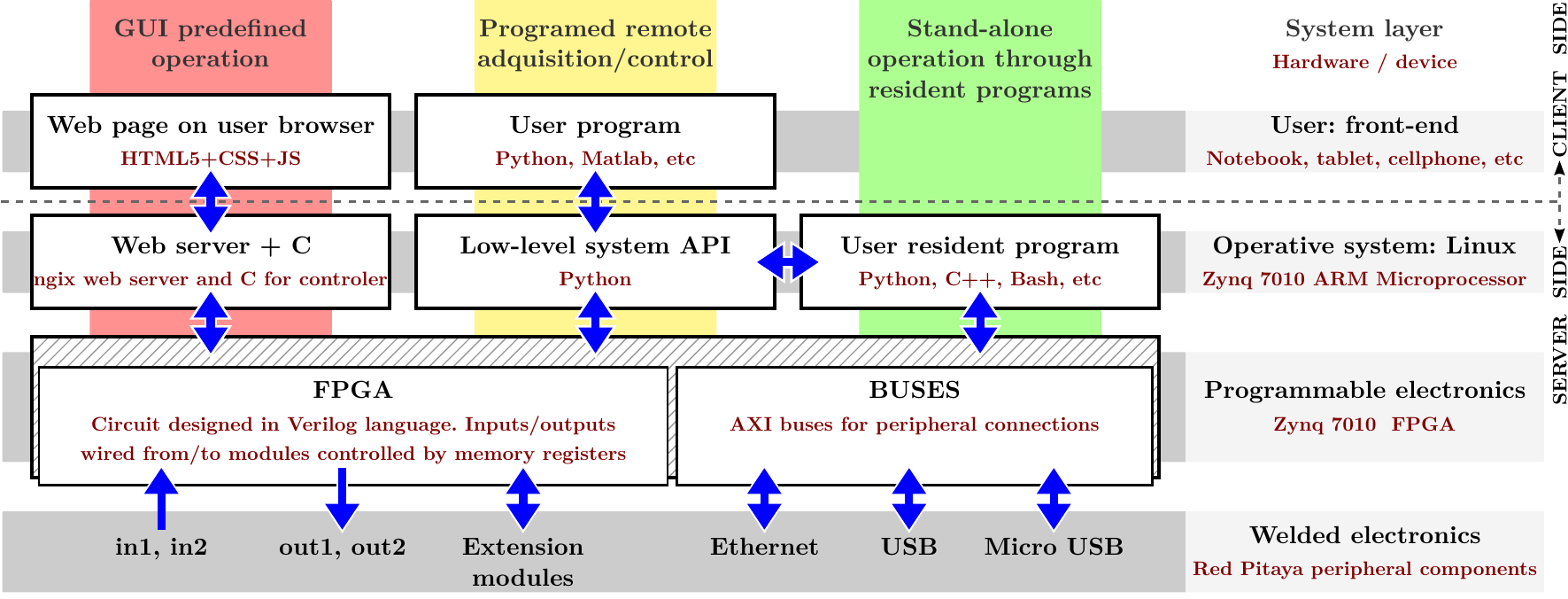, angle=0, width=0.95\textwidth}
  \end{center}
  \vspace{-15pt}
  \caption{\label{fig:app_design} Layer structured design for the client-server architecture of the toolkit.
            The elements are grouped horizontally by site and type. In columns, the user possible strategies
            are shown. The first column follows the Red Pitaya STEMLab original design philosophy. The dashed
            line separates the client part (end-user device) and the server part (STEMLab device). The
            arrows show communication direction between elements.   }
\end{figure*}

The advent of fast microprocessor devices and field programmable gate arrays (FPGA) has paved
the way to the implementation of the digital processing feedback schemes in experiment control setups. For example,
PIDs filters were implemented on pure-FPGA boards~\cite{RSI_Schwettmann_2011}. Also pre-processing information
like phase detectors was also accomplished with FPGAs~\cite{RSI_Xu_2012}.  A pure microprocessor implementation
of the feedback controller
has been showed to be more versatile and less complex in some simple or slow tasks~\cite{RSI_Huang_2014}.
Most of the experiences show the need to get a compromise between the fast and robust FPGA processing and the
versatile and easily re programmable microprocessor global functions and control.  This way, the FPGA can handle
time sensitive functions often also in parallel, while the microprocessor provides the user higher level functions
and control in an on-site configurable manner.

Embedded devices with FPGA and microprocessor were successfully used to achieve similar tasks as we show here ~\cite{RSI_Jorgensen_2016, RSI_Rico_2017}. However they were developed in the framework of proprietary,
platform-dependent software that limits the available tools for developing custom solutions and modifications.
An open-source system and open-hardware solution has also been presented with servo and loop-back control system
in pure FPGA for AMO~\cite{RSI_Leibrandt_2015}.

In this work we present a versatile general purpose toolkit for system control instrumentation built on a FPGA and
microprocessor embedded device that is ready for simple out-of-the-shelve usage and also for high complex integrated
schemes of acquisition and control.
The hardware is based on Red Pitaya STEMlab~125-14\cite{redpitaya_STEMLab} board,
which makes it a compact solution with a credit card size that can be acquired on web store.
The FPGA and software design is all open-source and can be accessed through a web repository\cite{ml_github}.
The toolkit includes oscilloscope, ramp/scan controller, two lock-in amplifiers, modulation generators, PID filters
and an overall control module which includes automatic locking, lock-watch and re-lock.
A big difference with previews implementations is the availability of a lock-in module which works with frequencies
up to 31 MHz. All stages of the high frequency lock-in are implemented in the Red Pitaya module without the use of
any extra components.

This device works as a stand-alone device that can be remotely controlled through Ethernet-TCP/IP network connection via a
web-based interface, and is therefore platform independent.  Moreover, one can log in
to the embedded operative system via SSH protocol and set operation parameters through a communication bus to the FPGA,
enabling different types of usage going from predefined easy configuration to low level user-defined programed control.

The present report is divided in two parts. In the first part we describe the toolkit hardware, software and usage.
We describe the user strategies options and the toolkit design in section~\ref{sec:user_strategy},
the architecture organized in layers in section~\ref{sec:layers} and the usage logic
organized in ``instruments'' in section~\ref{sec:instruments}.

In the second part, the toolkit is tested in several example applications in
an atomic, molecular and optical (AMO) physics laboratory.
We show a Rubidium absorption spectroscopy scheme in section~\ref{sec:Rb-abs},
with a Vertical-Cavity Surface-Emitting Laser
(VCSEL), using the toolkit for acquisition of transmission signal
and for side-fringe locking stabilization.
Then we use an External Cavity Diode Laser (ECDL) to implement
a saturated absorption spectroscopy scheme, in section~\ref{sec:satabs},
using lock-in phase sensitive acquisition to lock laser emission frequency to spectral peak maximum.
Finally, we present a Pound–Drever–Hall locking scheme for another ECDL, in section~\ref{sec:PDH}. One of the goals
of this proposal is the lock-in operation at 31~MHz, enabling the implementation
of the complete lock-in modulation and demodulation scheme in only one device,
something which has never been reported so far.

\section{System Design} 
\label{part:system_desing}

The toolkit is based on a commercial system on chip (SOC) device known as Red Pitaya v1.1 (STEMLab 125-14)
with a Xilinx Zynq 7010 integrated circuit core, that includes FPGA and a dual core ARM Cortex-A9 Processor,
integrated to peripherals suitable for signal acquisition and generation (fast ADCs, DACs, DIOs, and communication ports).
The device operation follows a client-server architecture that can be thought as a layer structure.
The client side is the user front-end running on clients own device. The server side is the STEMLab
device, including Operative System, programmable electronics and hardware peripherals.
The layers can be controlled on different levels depending on the user's requirements. In the following we describe
possible application strategies and the layer architecture, both depicted in figure~\ref{fig:app_design}.

\subsection{User Strategy}
\label{sec:user_strategy}

We present here three different user strategies with growing complexity and versatility. Depending on the needs,
an user can choose from simple pre-defined working modes to modifying instrument cabling or programming complex measurement
and feedback routines.

The ``GUI predefined operation'' option allows to perform most of the tasks here described from a web browser in the user device.
The front-end is part of the client-side and is built over HTML+JS page and presents a friendly
interface for several instruments operation: two lock-in amplifiers, two PID loops with different locking and re-locking routines,
an on-screen oscilloscope and cabling between them. These allow to perform a wide variety of measurement and control tasks which
we describe in detail in part~\ref{part:experiments},
with several examples which include low and high frequency lock-in measurements and active stabilization of lasers.

An even greater versatility can be achieved via the ``programed and remote acquisition and control'' option.
All the instruments can be remote controlled and data can be acquired trough user commands,
enabling the possibility of incorporating them to algorithms designed by the user in various programming languages.
Example code can be found in the project repository\cite{ml_github}
for Python and Matlab, including two channels oscilloscope acquisition,
on-demand multichannel reading of several signal values and instruments operation by writing control register values.

The user commands are Python scripts executed in Red Pitaya shell that implement the basic procedures for
reading and writing RAM addresses values linked to FPGA registers. The remote execution can be done by
any remote-shell tool, using serial communication (microUSB) or Ethernet (RJ45 port).
The examples provided on-line\cite{ml_github} use SSH protocol over TCP/IP network, being a secure and versatile
option that can be incorporated to almost any already working computer network. Moreover, open-source
implementations of SSH protocol are available for all possible operative systems of the client, enabling
the operation from any device or programing language.

The on-demand read and write procedures are only limited by network communication delays and microprocessor
process priority. The provided examples includes combined operation of instruments for tasks like
ramp/scan configuration, lock-in acquisition of system response, PIDs configuration and launching a triggered controlled
stabilization scheme.

We also provide an API for Python to simplify code writing of user designed  ``resident programs'', running in server side.
This enables the operation of instruments with lower latency and enables the design of algorithms for fast decision making,
faster multichannel acquisition and stand-alone working without client side. In this way, data acquisition and
instruments commands can be made with a latency of $\sim$ms order.

The ``resident programs'' advantages can exploited to implement different integration strategies of the toolkit in the lab.
For example, acquired data can be pre-processed on server-side and stored in the STEMLab device for later user retrieval.
On the other hand, the raw or the pre-processed data can be streamed to client side for on-line monitoring and storage.
The user can choose these or other options, depending on whatever suits the experimental scheme better.
In subsection\ref{subsec:allan} we report an example of long accumulated measurements streamed to the client-side,
used for Allan-deviation calculation.

\subsection{Layer architecture}
\label{sec:layers}

The system design can be understood in a 4 layer scheme (see figure~\ref{fig:app_design}). The lower three layers are part
of the server side and take place in different parts of STEMLab board hardware.

The lowest one is the ``Welded electronics layer'', that provides the connection to digital and analog input/output ports
and the access to peripherals, including some analog electronics for signal conditioning.
Signals in this layer are converted from continuous analog voltage signals on wires to digital
clocked buses of data, suitable for FPGA reading.
For this purpose the board includes two fast Analog to Digital Converters (ADC), associated with
\texttt{in1} and \texttt{in2} FPGA buses, and two fast Digital to Analog
Converters (DAC), associated with \texttt{out1} and \texttt{out2} FPGA buses, working at 125~MSa/s
with 14 bits of resolution, on the range of~$\pm1\,\text{V}$. ADC inputs can work in the range of~$\pm10\,\text{V}$
with hardware jumper configuration.
Also, 16 input/output digital pins are included, with a maximum refresh rate of 125~MSa/s, on a LVCMOS 3.3~V logic.
It also includes four slow outputs of 0--1.8~V, 1~MSa/s build from low pass filtered PWM signals, and four slow inputs
with the same sample rate.

The ``Programmable electronics layer'' stores the digital electronic FPGA circuits for real-time processing,
like filters, mixers, adders, etc, that build up the core of each instrument implementation.
In this layer the signals data flow through digital buses that represent signed integer values of 14 bits resolution.
The circuits were designed in Verilog language, synthesized using Xilinx Vivado 2005.2 software and implemented on the Zynq-7010 FPGA chip on
the Red Pitaya board. The Xilinx development tools for this chip are available for free, so no additional costs are incorporated on
development licenses if the design should be modified.
This layer also provides connectivity between the microprocessor and ``Welded electronics layer'' through specialized data buses.

The FPGA design was made with simplicity in mind, to ease the user understanding of the electronic logic. Also, the
implementation was made trying to prioritize direct wire processing, reducing the usage of registers in the middle
of input-output signal flow. This decision was taken because each register adds a clock period delay (8~ns) to the data flow.
Closed-loop control schemes are bandwidth limited by this delay value\cite{RMP_Bechhoefer_2005}. The achieved delay for
device input-output using one PID filter was 130~ns, which imposes a theoretical limit to the feedback bandwidth of
3.8 MHz.

The ``Operative system layer'' stores the back-end logic that controls the operation of the digital circuits.
It runs a GNU/Linux operating system with a set of RAM memory addresses mapped to FPGA registers that are used by the
instruments circuits to set their configurations or store data.
The back-end logic reads and writes these registers and makes some data conditioning, like conversion from FPGA integer-raw
data to end-user float-voltage data values.
This logic is implemented in the API (programmed in Python) and also in the Nginx web server
(programmed in a C extension module). The web server provides the client access and control from user web browser,
exchanging data using JSON standard format and HTTP/POST protocol. The API is used by the user commands introduced
in the section~\ref{sec:user_strategy}.

The ``User Front-end'' layer is only provided for the ``GUI predefined operation'' strategy, and it consists of a dynamic web page loaded in the client device.
The data acquired in FPGA layer, and conditioned in the back-end layer, is
shown here in a intuitive way, as can be seen in~\ref{fig:web_gui}

\subsection{The instruments} 
\label{sec:instruments}

The functionalities of the toolkit have been organized in logical units
called ``instruments'', that allows the user to interact with
it. Each instrument provides some of the functionalities of typical laboratory
equipments used for acquisition and control. The  instruments implemented are:

\begin{enumerate}
  \item Two Lock-in amplifiers
  \item Two Proportional-Integral-Derivative (PID) controllers
  \item A Ramp/Scan function generator
  \item A general Lock-control helper, with event monitor and re-lock routine
  \item An Oscilloscope
\end{enumerate}

Each instrument is composed by a set of HTML controls in the
Web GUI, grouped in panels, and a set of Verilog modules for the FPGA layer.
Instruments 2 and 5 are modified versions of the open-source applications developed by Red Pitaya community
and released with a BSD license\cite{redpitaya2015}.
Instruments 1, 3 and 4 were developed for this work.

In the FPGA layer, the instruments are composed by logical modules whose behavior is controlled and monitored by registers.
The modules are independent interconnected circuits which handle signal acquisition, processing and
generation. They also include internal memory which allows implementing filters and control loops.
The base modules include: function generators, ramp generator, multipliers, demodulators, low-pass filters, multiplexers, etc.

The control registers can be set and read from Red Pitaya local shell, from remote software or from
the aplication Web GUI. The last one includes user friendly options for operation, described with
more detail in section~\ref{sec:Web-GUI}.

A set of buses and multiplexers allow the interconnection of the different instruments.
The buses transport the ADC input signals and the
instruments outputs signals. The multiplexers are controlled by the user through register
values to chose the instruments inputs or the DAC outputs signals.
In this way, several instruments can be combined into a system with dedicated purpose,
such as lock-in demodulation and PID control for stabilizations schemes.

Core design of the instruments is described in this section and usage information
is documented in the project web page\cite{ml_github_doc}.

\subsubsection{The lock-in amplifiers}
\label{subsec:lock-in_intrument}
Two lock-in amplifiers were implemented to cover
different kinds of applications: a ``standard'' harmonic lock-in, with a frequency range that
goes from 3~Hz to 49.6~kHz, and a square wave lock-in with wider frequency range,
from 30~mHz to 31~MHz.

\begin{figure}[ht]
  \begin{center}
    \epsfig{file=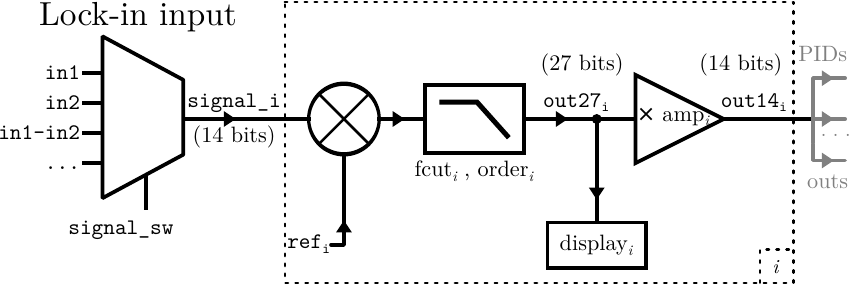, angle=0, width=\columnwidth}
  \end{center}
  \vspace{-15pt}
  \caption{\label{fig:fpga_scheme_lock-in} FPGA designed logic for the lock-in demodulation paths.
            The Lock-in input is a signal selector common to all the demodulation paths: \texttt{signal\_i}.
            The part inside the dashed line box is repeated 7 times, changing the \texttt{ref\_i} reference signal
            and the elements parameter ranges.}
\end{figure}

Figure~\ref{fig:fpga_scheme_lock-in} shows the scheme of the FPGA implementation logic of a lock-in
demodulation path. The input signal \texttt{signal\_i} is multiplied by a local oscillator
$\texttt{ref}_\texttt{i}$ signal that settles the frequency and phase reference,
and then is filtered by a low pass filter (LPW) with a cut-off frequency of
$\texttt{fcut}_\texttt{i}$. The LPF output $\texttt{out27}_\texttt{i}$ is a 27~bit signed int
register which is available for recoding measurements and which can be monitored via the web GUI.
A bus-trim applied to the register reduces it to a 14~bits signal $\texttt{out14}_\texttt{i}$.
The selection of the trimmed bits has the net effect of an amplifier by powers of 2, controlled
by the $\texttt{amp}_\texttt{i}$ parameter.
This 14 bit singal can then be connected to either the PIDs, the output DACs or the oscilloscope
input channels.

\begin{table}
  \caption{\label{tab:lock-in} Reference of lock-in demodulation paths depicted in figure~\ref{fig:fpga_scheme_lock-in}.
                               The 1--5 paths are part of the
                               harmonic lock-in, and the 6--8 are part of the square wave lock-in.}
  \small
  \begin{ruledtabular}
    \begin{tabular}{lllll}
      \texttt{i}
      & \multicolumn{1}{p{1.6cm}}{\raggedright \texttt{ref} \\ name }
      & \multicolumn{1}{p{1cm}}{\raggedright signal \\ equation }
      & \multicolumn{1}{p{1cm}}{\raggedright \texttt{out27} \\ name }
      & \multicolumn{1}{p{1cm}}{\raggedright \texttt{out14} \\ name  }    \\
      \hline
       1    &  \texttt{cos\_ref}& $\cos(2\pi\;\phantom{2}f \,t)$                 & \texttt{X}   & \texttt{Xo}   \\
       2    &  \texttt{sin\_ref}& $\sin(2\pi\;\phantom{2}f \,t)$                 & \texttt{Y}   & \texttt{Yo}   \\
       3    &  \texttt{cos1f}   & $\cos(2\pi\;\phantom{2}f \,t - \phi)$          & \texttt{F1}  & \texttt{F1o}  \\
       4    &  \texttt{cos2f}   & $\cos(2\pi\;2f\,t - 2\phi)$                    & \texttt{F2}  & \texttt{F2o}  \\
       5    &  \texttt{cos3f}   & $\cos(2\pi\;3f\,t - 3\phi)$                    & \texttt{F3}  & \texttt{F3o}  \\
       \hline
       6    &  \texttt{sq\_ref} & $\text{sgn}(\cos(2\pi\;f_{sq} \,t          ))$ & \texttt{sqX} & \texttt{sqXo}   \\
       7    &  \texttt{sq\_quad}& $\text{sgn}(\sin(2\pi\;f_{sq} \,t          ))$ & \texttt{sqY} & \texttt{sqYo}   \\
       8    &  \texttt{sq\_phas}& $\text{sgn}(\cos(2\pi\;f_{sq} \,t - \varphi))$ & \texttt{sqF} & \texttt{sqFo}   \\
    \end{tabular}
  \end{ruledtabular}
\end{table}

The Harmonic Lock-in is composed of five demodulation paths like the depicted in figure~\ref{fig:fpga_scheme_lock-in}.
Each path has its own $\texttt{ref}_\texttt{i}$ signal and buses names, showed in table~\ref{tab:lock-in}, that
can be used in the application for DAC outputs or oscilloscope visualization.
The \texttt{cos\_ref} and \texttt{sin\_ref} paths have their reference in quadrature, so the \texttt{X} and
\texttt{Y} outputs provide the full phase and amplitude information of the \texttt{signal\_i} filtered frequency component.
The other three paths allow to set a fixed phase relation respec to \texttt{cos\_ref} and
also obtain information about the first ($2f$) and second ($3f$) harmonics.
The $\phi$ parameter, which can be configured through the Web GUI \texttt{phase} control,
sets a phase displacement of
$\phi = 2 \pi \frac{\texttt{phase}}{2520}$.

Each local oscillator signal is made from 2520 signed integers values proportional to a sine
period. They are stored in a memory module implemented as various lookup tables.
The reading addresses for the memories modules are set by counter modules, driven by a
configurable clock divider that allows to change the reading sample rate and, whit it,
the resulting $\texttt{ref}_\texttt{i}$ working frequency.
The \texttt{cos\_ref}, \texttt{sin\_ref}, \texttt{cos2f} and \texttt{cos3f} signals
were designed to satisfy the discrete Fourier orthogonality relations of equation~\eqref{eq:fourier},
which avoid offsets generated by digital rounding, and is important not only for measurement precision but also
useful for reducing instrumentation induced instabilities in a feedback stabilization scheme.
This condition is not satisfied if we use signals built from real valued $cos$ function discretized
with a global criteria applied to all the values (like using standard integer conversion functions:
\texttt{floor}, \texttt{ceil} or \texttt{round}).

\begin{equation} \label{eq:fourier}
  \sum_{i=0}^{2519} { \text{\texttt{f[i]}} \cdot \text{\texttt{g[i]}} } = 0
  \;\;
  \text{ for } \texttt{f} \neq \texttt{g}
  \;\;
  \text{ and }
  \texttt{f},\texttt{g} =
  \left[ {
    \begin{array}{c}
      \texttt{cos}    \\
      \texttt{sin}    \\
      \texttt{cos2f}  \\
      \texttt{cos3f}
    \end{array}
  } \right]
\end{equation}

The square wave lock-in is composed of three demodulation paths: 6-8 from table~\ref{tab:lock-in}. Two local
oscillators in quadrature (\texttt{sq\_ref} and \texttt{sq\_quad}) allow the acquisition of the
complete quadrature information in the \texttt{sqX} and \texttt{sqY} registers.
Also, another
oscillator path, \texttt{sq\_phas}, with configurable phase relation is available, with output value stored
in \texttt{sqF}.

The square wave signals are build on run-time, switching a bit that represents the $\pm1$ values. The
working frequency $f_{sq}$ is set by defining the semi-period time length, counting steps the base FPGA
clock. In this way, the period can be set with a resolution of 8 ns, starting from 32 ns (31.25 MHz)
and expanding the possible values up to 68 seconds. The $\varphi$ phase relation is set as an integer factor
of 8~ns. The multipliers of these paths are modified to map the binary input from 0,1
values to $\pm1$. The \texttt{sq\_ref} binary signal is available in one of the fast binary outputs of
the extensions pins of STEMLab device, and all the signals can be used on the DACs outputs, where they are mapped
from the binary values to $\pm0.5$ volts.

The \texttt{out27} resgister enables lock-in measurements with a full resolution of 27 bits and a sensitivity
of $59.6\,\mu\text{V}$. The sensitivity can be enhanced with pre-amplification, as is shown in experience of
section~\ref{sec:satabs}.

\subsubsection{PID controllers}
\label{subsec:PID_intrument}

Two PID modules were implemented for use in feedback stabilization schemes.
The circuit design is based on the PID included in the free software applications of
Red Pitaya community\cite{redpitaya2015}.
They where modified to extend the working parameters range over several orders of magnitude, by
incorporating scale registers.
The output control was enhanced by adding
features like output value freezing and integrator memory freezing, useful for re-lock routines (see \ref{subsec:lock-control_intrument}).

\begin{figure}[ht]
  \begin{center}
    \epsfig{file=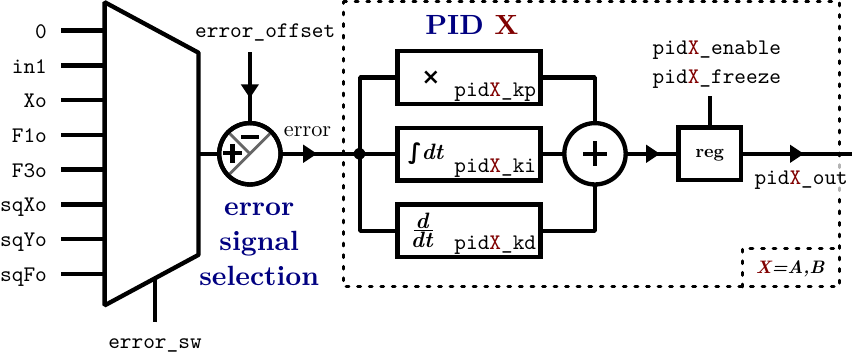, angle=0, width=\columnwidth}
  \end{center}
  \vspace{-15pt}
  \caption{\label{fig:fpga_scheme_PIDs} FPGA designed logic for PID modules.
            The error signal selection lets choose a common input \texttt{error} fro both PIDs.
            The elements inside the dashed line box are replicated once for each PID.
            Other inputs can be selected independently for each PID, but is not included
            in this graphic.}
\end{figure}

Both modules have a common \texttt{error} signal as default input, that can be selected from different
input buses, as is shown in figure~\ref{fig:fpga_scheme_PIDs}.
The input can be shifted by a user-controlled \texttt{error\_offset} value that can be interpreted as a
working setpoint.
Also, independent input signals for each module can be chosen (not shown in the figure).

Each module implements three filters, proportional, integral and derivative, that sums up to build the
output signal \texttt{pidX\_out} ($\texttt{X}=\texttt{A},\texttt{B}$), as it is showed in
figure~\ref{fig:fpga_scheme_PIDs}. The summed result is stored in a register which allows to freeze the
value at command.

The behavior of the three filters depends on the value of three parameters: $k_p$, $k_i={}^1/_{\tau_i}$
and $k_d=\tau_d$. Two registers allow the user to control each of them: one 14~bit signed int
value changes the parameter linearly  while the second value allows to change the order of magnitude
of the parameter, working as a scale control.
For example,
the $k_p$ parameter is controlled by the \texttt{pidX\_kp} register value, and by the scale factor
set by the register \texttt{pidX\_PSR}, that allows to choose uno of the predefined values for
$n_p$ in equation~\eqref{eq:pid_kp}. The same logic is applied to control $k_i$,$k_d$ using
\texttt{pidX\_ki},\texttt{pidX\_kd} and \texttt{pidX\_ISR},\texttt{pidX\_DSR} respectively, as is shown
in equations~\eqref{eq:pid_ki} and~\eqref{eq:pid_kd}:

\begin{subequations}
\begin{align} \label{eq:pid_kp}
  k_p & = \frac{ \texttt{pidX\_kp} } { 2^{n_p} }
  & \text{ with }
  & {n_p} = \{ 0, 3, 6, 10, 12 \} \\[2mm]
  \label{eq:pid_ki}
  \tau_i & = \frac{ 2^{n_i} \cdot 8 \, \text{ns} } { \texttt{pidX\_ki} }
  & \text{ with }
  &  {n_i} =
  \begin{array}{l}
     \{0, 3, 6,10,13,\\
    \;\;16,20,23,26,30 \}
  \end{array} \\[2mm]
  \label{eq:pid_kd}
  \tau_d &  = \frac{ 2^{n_d} \cdot 8 \, \text{ns} } { \texttt{pidX\_kd} } \cdot \frac{60}{64}
  & \text{ with }
  &  {n_d} = \{ 0, 3, 6,10,13,16 \}
\end{align}
\end{subequations}

The linear coefficient register lets the user have fine control over the PID parameters,
while the power of two factor, implemented efficiently through shift registers, expand
the parameters scope over several orders of magnitude.
This dual scale allows controlling systems with time constants ranging from a few $\mu$s to many seconds.

The derivative module of the PIDs was completely rewritten to avoid high frequency noise amplification.
The new design ensures that spikes and frequency components whose characteristic times are below the order of magnitude
settled by the $n_d$ value won't induce undesired perturbations that can make the feedback scheme unstable.
The implemented design includes a down-sampling procedure and a slope calculation sub-module, called \texttt{slope9}.
The PID input signal is down-sampled taking the mean value of $2^{n_d}$ consecutive data samples, so the processed
signal that feeds the \texttt{slope9} submodule has a ladder shape with a $T_d = 9 \cdot 2^{n_d} \cdot 8 \text{ns}$
step time length. The net effect of this procedure is similar to the application of a low pass filter
with time constant of $T_d$ before the derivative calculation.
The \texttt{slope9} submodule calculates the signal derivative by taking the slope of a linear square least regression
of the last nine steps values of the down-sampled signal.

The parameters $k_p$, $\tau_i$ and $\tau_d$ can be used for prediction of an ideal PID response to
an $e(t)$ input signal using the equation~\eqref{eq:pid_cont}.
The equation~\eqref{eq:pid_discrete} provides a more accurate prediction of the implemented PID
response in clock steps of 8~ns. The ``slope9out'' function represents the \texttt{slope9} submodule output signal
that can be simulated by an algorithm following the procedure described above.

\begin{subequations}
\label{eq:pid_response}
  \begin{align} \label{eq:pid_cont}
    \texttt{pid}_\texttt{out}(t) &=
                k_p              \cdot e(t) \; +\;
                \frac{1}{\tau_i} \cdot \int_0^t e(t') \, d{t'} \;+\;
                \tau_d           \cdot \frac{d}{dt} e(t)
    \\[2mm] \label{eq:pid_discrete}
    \texttt{pid}_\texttt{out}[n+1] &=
                k_p                         \cdot e[n] \; +\;
                \frac{8\,\text{ns}}{\tau_i} \cdot \sum_{i=0}^n e[i] \;
                \nonumber \\  & \; \; \; \; +\;
                \frac{\tau_d}{8\,\text{ns}} \cdot \text{slope9out}( e[n,\ldots,n-9\cdot 2^{n_d}+1 ]  )
  \end{align}
\end{subequations}

\subsubsection{Ramp function generator for Scanning}

A triangle wave-shape generator was implemented for scanning purposes.
Two synchronized signals are built on run-time, \texttt{ramp\_A} and \texttt{ramp\_B},
useful to control two system parameters at the same time.

\begin{figure}[ht]
  \begin{center}
    \epsfig{file=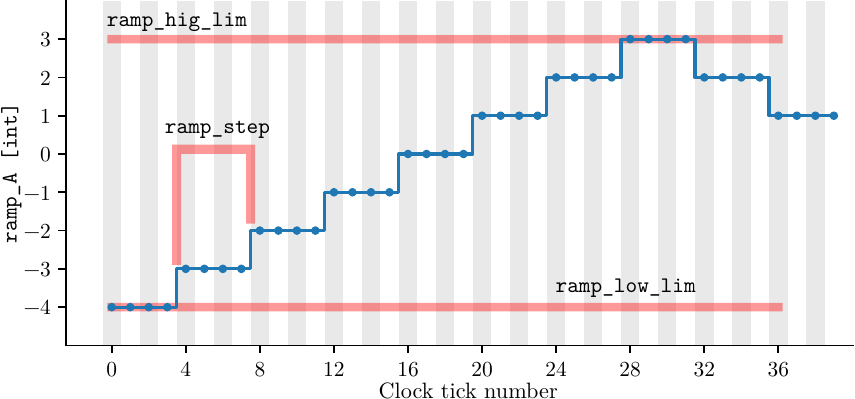, angle=0, width=\columnwidth}
  \end{center}
  \vspace{-15pt}
  \caption{\label{fig:fpga_scheme_ramp} \texttt{ramp\_A} behaviour os Ramp instrument
           for sample parameters: \texttt{ramp\_hig\_lim}, \texttt{ramp\_low\_lim}, \texttt{ramp\_step}.
           The clock tick period id 8~ns.}
\end{figure}

The \texttt{ramp\_A} behavior is controlled by setting the triangle sweep range and the
time length $T$ of each step of the ladder-shaped signal (see figure~\ref{fig:fpga_scheme_ramp}). The registers
\texttt{ramp\_hig\_lim} and \texttt{ramp\_low\_lim} control the top and bottom limits, and
the \texttt{ramp\_step} register sets $T = \texttt{ramp\_step} \cdot 8\,\text{ns}$,
all of them accessible from Web GUI\@. After each $T$ time period the \texttt{ramp\_A}
value increase/decrease by 1~int until it reaches one of the limits, and then changes the slope sign.
\texttt{ramp\_B} is generated multiplying \texttt{ramp\_A} by the factor \texttt{ramp\_B\_factor}/4096~.

The instrument allows the user to select the \texttt{ramp} slope and limits. Since the acquired signal waveform may change with sweeping
speed, affected by bandwidth limits of intermediate elements, the slope is kept constant
when the range limits are changed by the user. This lets the user see the same waveform
through all the process. This design has been helpful when using the \texttt{ramp} for
spectroscopy acquisition schemes, making it easier to choose the scanning range.

The ramp generator can be sarted and stopped in any time through the \texttt{ramp\_enable} register,
can be reseted to \texttt{ramp\_A}=0 value, and the starting slope sign can be set.
Also, the Lock-control module can take control
of the start/stop command and the ramp limits for locking and re-locking procedures.

\subsubsection{Lock-control}
\label{subsec:lock-control_intrument}

\begin{figure}[ht]
  \begin{center}
    \epsfig{file=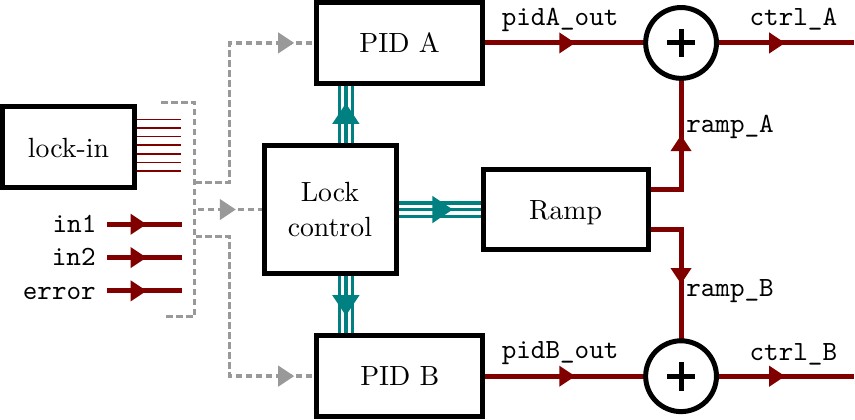, angle=0, width=\columnwidth}
  \end{center}
  \vspace{-15pt}
  \caption{\label{fig:fpga_scheme_lock_ctrl} Lock-control links to PIDs and Ramp instruments.
            \textcolor{darkred}{Red lines} are signal data buses and \textcolor{turquoise}{turquoise lines} represents sets of buses for
            parameters configuration. All the Lock-in outputs and the ADC inputs can be used
            in the PIDs for signal processing and in the Lock-control for event identification and trigger.}
\end{figure}

A device meant for locking to a given \texttt{error} signal normally requires switching between different behaviors:
scanning, locking, re-locking on events, etc. Changing from one state to the other requires timed interaction
between the previous modules. This interaction is handled by the Lock-control module.
Figure~\ref{fig:fpga_scheme_lock_ctrl} shows the schematic inter-cabling of the modules.

A typical example of this behavior is what we call \textit{Trigger Lock}. Here the system switches from a
scan-mode, where one can see the error signal by sweeping the control parameter, to a lock-mode where the
scan is turned off and the PIDs are turned on to stabilize the error signal.
That is, on a trigger event, the ramp's outputs are frozen and the PID's outputs are enabled.
The trigger can be set to a given value of the ramp period (\textit{time trigger}), to a given value of the signal
(\textit{level trigger}) or the fulfillment of both conditions (\textit{level+time trigger}). The values for
these triggers can be selected graphically on the Oscilloscope screen or manually settled.

The module also provides a \textit{Re-lock} routine which tries to automatically re-lock on eligible events.
This submodule will consider the system is out-of-lock when the absolute value of the \texttt{error} signal
exceeds a set value or when one of the inputs (\texttt{in1}, \texttt{in2}, \texttt{in1}$-$\texttt{in2}, or any
lock-in demodulated signals)
drops below a set threshold.
When any of these conditions is met the re-locking routine is started.
The re-locking procedure was inspired on a recent work\cite{RSI_Leibrandt_2015} and
consists on freezing the PIDs and starting a ramp scan
increasing the scan limits on a two factor in each half period.
If during scanning the system reaches a lock condition, the submodule stops the scan and turns on the PIDs again.
If the lock condition is never met, the submodule will continue increasing the scan amplitude
until it makes a semi-period sweep with the longest width (1~V $\equiv$ 8192 int),
and then will stop ramp.

Finally, a submodule which carries out a \textit{Step Response Measurement} allows the user to evaluate the
system response to an abrupt change in the control value. This information can then be used to calibrate the
parameters of the PID modules.

\subsubsection{Oscilloscope}
\label{subsec:scope_intrument}
The Oscilloscope instrument is a modified version of Scope Application from Free Software Red Pitaya
developers community\cite{redpitaya2015}.
It's composed of two memory arrays with a length of 16384
for storing  14~bit signed int values, modules for triggering,  decimation control and anti-aliasing filters, and it allows to make acquisitions on FPGA clock sample rate (125 MSa/s).
Here we extended the functionality of the basic oscilloscope to allow the display of various internal and external signals.
These include the
ADC inputs (\texttt{in1},\texttt{in2}), instruments outputs
(PIDs and Ramp), important link buses (\texttt{error}, \texttt{ctrl\_A}, \texttt{ctrl\_B}) and all the
lock-in paths outputs and local oscillators.
With this extension, the oscilloscope is useful not only for
external data acquisition but also for internal data flow and processing inspection, essential for debugging and
tunning the parameters of each module.

Also, some acquisition options where added. We included the possibility to disable the anti-aliasing FPGA module filter
and to incorporate more triggers from useful signals. The external trigger is extended for user selectable event list,
including out-of-lock, triggered lock-start, \texttt{ramp\_A} at limit reach and lock-in oscillators period start.
Also, some web GUI options were added, like an R-$\phi$ (absolute and phase) run-time calculation option for lock-in
\texttt{X},\texttt{Y} and \texttt{sqX},\texttt{sqY} visualization and switch between Volts and int units.
By default, the acquired curve values are expressed in Volts, using the ADC/DAC conversion factor (signed 14~bits int resolution):
$8192\,\text{int}\equiv 1\,\text{V}$. This can be switched to raw int values.

\subsection{The Web GUI} 
\label{sec:Web-GUI}

The Web GUI frontend design is also based in free software Scope application\cite{redpitaya2015}.
It provides the out-of-the-shelve functionality showed in the first column of figure~\ref{fig:app_design}.
It is structured in columns with dropdown panels that groups configurations settings by functionality.
We improved the interface adding several useful functions, like data save/load, configuration save/load, stop
button, etc.

\begin{figure*}[ht]
  \begin{center}
    \epsfig{file=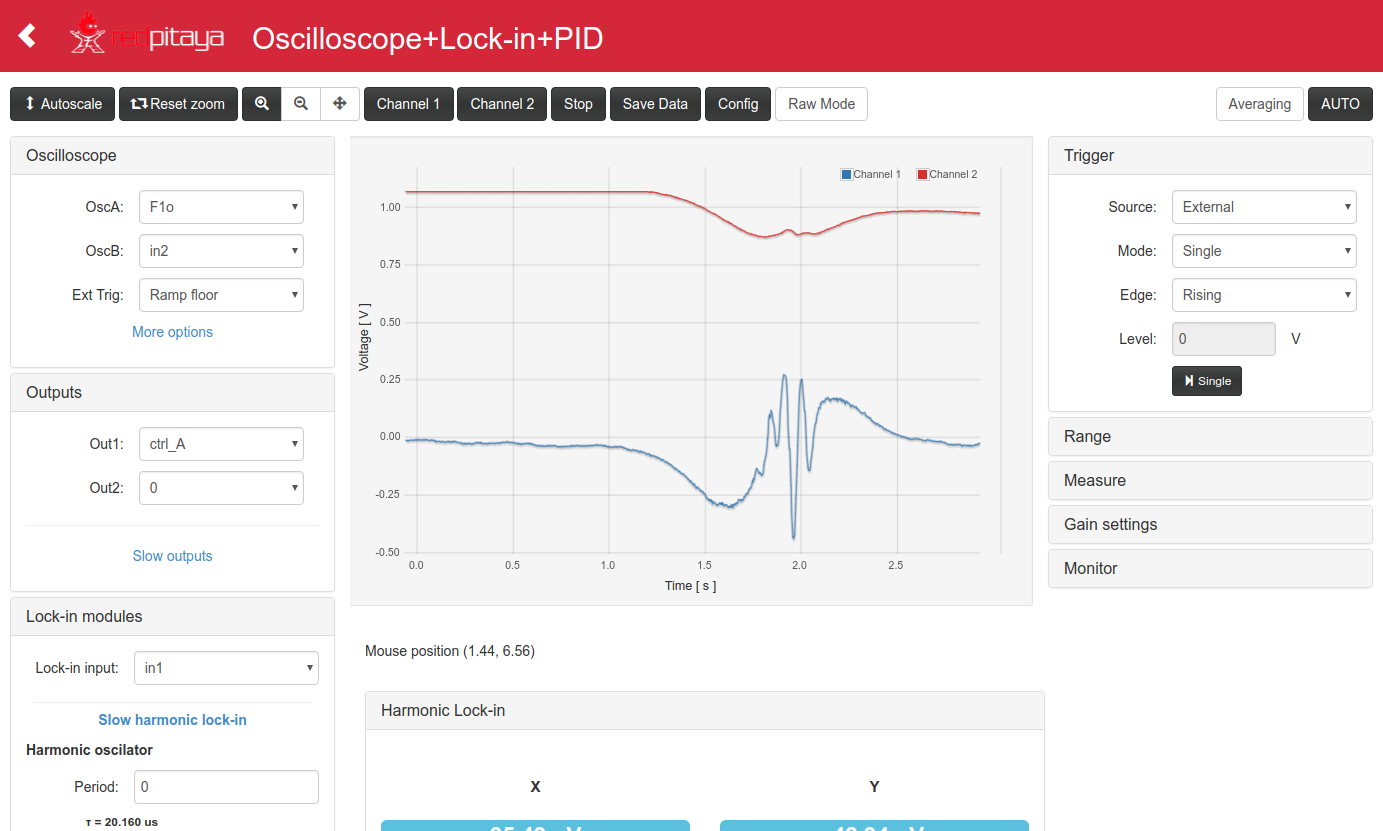, angle=0, width=\textwidth}
  \end{center}
  \vspace{-15pt}
  \caption{\label{fig:web_gui} Web page GUI frontend, organized in three columns and panels. The left one
            contains one panel per instrument with the configurations options for each one. The right one
            contains the oscilloscope visualization options. The middle one contains the oscilloscope
            screen and panels for \texttt{out27} lock-in measured values visualization.}
\end{figure*}

A left column was added for placing the instruments panels designed for this work, and
two bottom panels for the lock-in amplifiers outputs visualization (figure~\ref{fig:web_gui}).
The instruments controls were designed with the philosophy of keeping simplicity without
compromising the low-level accuracy. Most of the controls use integers numbers for
input data type, that allows the user define the precise value of the FPGA registers
that the circuits will use, and includes text displays that translates this
values to the physical magnitudes involved (i.e., seconds, volts).

The oscilloscope screen is in the middle of the page. It displays two channels for plotting user selected signals.
In the sample applications discussed below, in part~\ref{part:experiments},
one channel shows the spectral response of the system while the other displays the \texttt{error} signal.
This provides a fast way to view and evaluate the performance of the system on lock, entering a
lock, and re-locking.


\section{Example applications} 
\label{part:experiments}
We report a set of experimental applications of the toolkit presented in order of complexity
to show the instruments usability and potential.
In the first one (section~\ref{sec:Rb-abs}) we introduce the usage of the Ramp and the Oscilloscope
to take the absorption spectrum of an atomic vapor, with a simple scheme of one control parameter and
one measured variable.
Also, we implement a side-of-fringe locking scheme using a PID controller an the Lock Control instrument.
In the second application (section~\ref{sec:satabs}) we add the Harmonic Lock-in amplifier
to the measure and stabilization scheme, in a saturated absorption experiment. We used two output ports
and measure two input variables, in a multi-input-output control system, but with only one control parameter.
In the third one (section~\ref{sec:PDH}) we show the Square Lock-in amplifier working at 31~MHz for
a Pound-Drever-Hall stabilization application. Some advanced capabilities are shown here, like multiple controlled
variables, in-device programing for special measurements and different hardware implementations of same stabilization scheme.
Finally, we discuss some potential applications of the toolkit.

\subsection{Absorption spectroscopy:\\* Ramp, Oscilloscope \& PIDs} 
\label{sec:Rb-abs}

\begin{figure}[ht]
  \begin{center}
    \epsfig{file=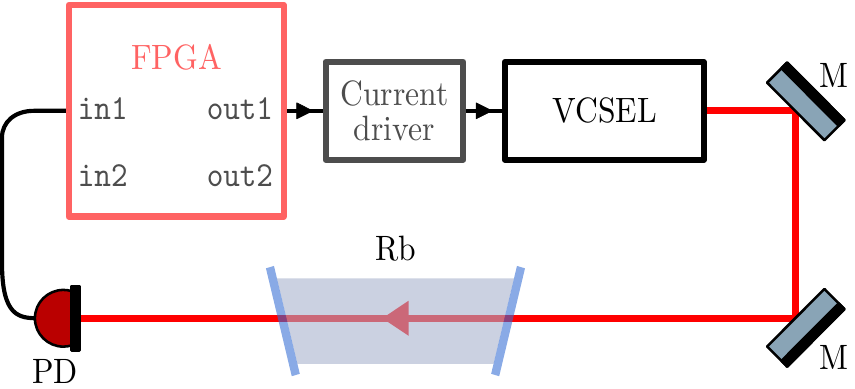, angle=0, width=\columnwidth}
  \end{center}
  \vspace{-15pt}
  \caption{\label{fig:abs-exp} Scheme of the Rubidium absorption spectroscopy experience.
            Only one system parameter is controlled (laser diode current) trough \texttt{out1} output,
            and one variable is sensed (photodiode response) trough \texttt{in1} input port.
            M: mirror; PD: photodiode; Rb: Rubidium vapor cell.}
\end{figure}

We tested the out-of-the-shelve capabilities of the toolkit in an atomic
vapor absorption spectroscopy experiment with Rubidium.
A tunable laser that goes across an atomic vapor cell is used to make a optic-frequency
scan around one of the atomic electronic transitions. A photodiode is used to measure the
intensity decay at the cell output (figure~\ref{fig:abs-exp}), as a result of the absorption
on transition resonance.
We used a VCSEL, what ensures
mode-hop-free scanning using only one control parameter: the diode current\cite{tunableVCSEL_2000}.
The laser wavelength is centered in 795~nm, suitable for Rubidium $\text{D}_1$ line
($5^2 \text{S}_{1/2} \longrightarrow 5^2 \text{P}_{1/2} $) spectroscopy.

The Ramp instrument was used to  make the frequency scan,
by performing a voltage-controlled sweep of the current driver.
The Oscilloscope instrument was used for photodiode signal acquisition.

\begin{figure}[hb]
  \begin{center}
    \epsfig{file=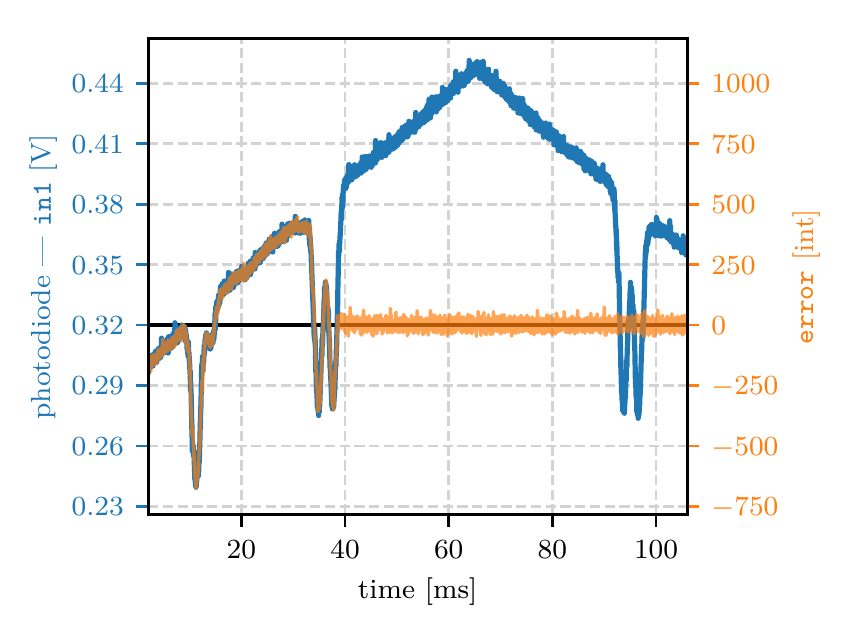, angle=0, width=\columnwidth}
  \end{center}
  \vspace{-15pt}
  \caption{\label{fig:abs-exp-data} Absorption spectroscopy of Rubidium.
            The transmitted intensity across the Rubidium cell is plotted
            for a complete Ramp scan measurement (\blue line) and for a
            lock-start event (\orange line). The curve dips corresponds to
            ${}^{87}\text{Rb}$ and ${}^{85}\text{Rb}$ hyperfine absorption lines.  }
\end{figure}

The Ramp was configured to make a scan of $\sim1\,\text{V}$ through \texttt{out1}
DAC at $7\,\text{Hz}$. In figure~\ref{fig:abs-exp-data} the
transmitted signal measured by the photodiode is shown with \blue line. The base slope of
the curve is related to the laser power increment along the scan because of the
current variation. The four dips in the curve are the absorption lines
for each of the hyperfine-split transitions.

This experiment was used to test the most simple stabilization scheme: side-fringe locking.
The variable to stabilize is the transmitted signal, measured from \texttt{in1} port.
The PID instrument was used to make the feedback response and the Lock Control instrument
was used to control loop-close event.

The locking set-point was configured on  $\texttt{error\_offset}=2620\,\text{int} \cong	0.32\,\text{V}$
to lock the side of the last spectrum dip in the Ramp scan, making $\texttt{error} = \texttt{in1} - 3300 $.
To prevent the lock on the first cross-over of  \texttt{error\_offset} the Lock-control instrument
was configured to close the loop on \textit{level+time} trigger.
The loop-close procedures was configured to stop the Ramp scan and enable the \texttt{pidA\_out}.
The PID parameters were set to $\tau_i\texttt{[ki=600]}\cong 895\,\mu s$ and $k_p\texttt{[kp=1024]} = 1$.

The result after reaching Lock-control \textit{level+time} trigger condition is shown in the figure~\ref{fig:abs-exp-data}
(\orange line). The transmitted signal remained in the configured set-point level with
a standard error of $\sigma = 1.85(3)\,\text{mV}$.

\subsection{Saturated absorption spectroscopy:\\* Harmonic Lock-in} 
\label{sec:satabs}

\begin{figure}[hb]
  \begin{center}
    \epsfig{file=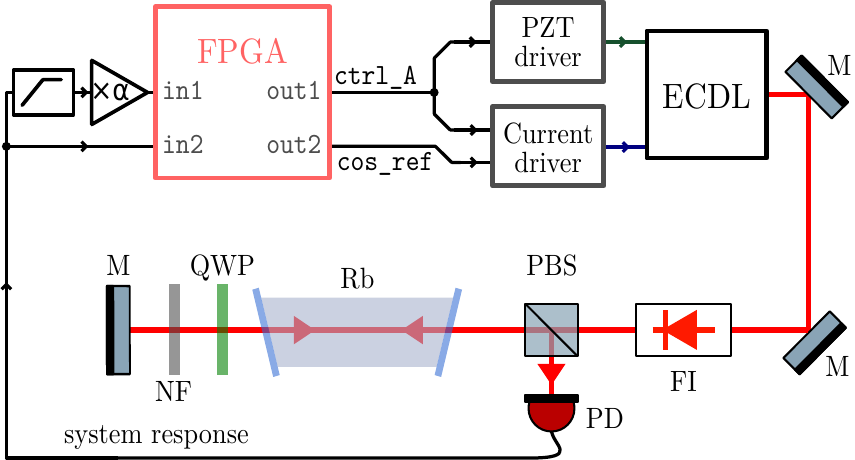, angle=0, width=0.5\textwidth}
  \end{center}
  \vspace{-15pt}
  \caption{\label{fig:SatAbs_schema} Scheme of saturated absorption spectroscopy of Rubidium.
            The ECDL emission wavelength is controlled trough two parameters: PZT voltage and diode current.
            A harmonic modulation is incorporated through current driver.
            The lase going from right to left is the pump beam. The reflection is the probe beam.
            Two polarization rotations with the  QWP and the PBS are used to recover the
            outgoing probe beam and measure it with the photodiode. Direct photodiode voltage and
            an AC amplified line are registered by the FPGA fast input potrs.
            M: mirror; QWP: quater wave plate; PBS: polarizing beamsplitter;
            FI: Faraday isolator; PD: photodiode; Rb: Rubidium vapor cell. }
\end{figure}

In this experience we extended the previous setup to make a Saturation Absorption
Spectroscopy\cite{SatAbs_Bhattacharyya2004} scheme, that allows
to measure transitions with resolution exceeding the limit posed by Doppler broadening\cite{SatAbs_Preston1996}.
This scheme uses a pump and probe configuration, implemented here with the same beam
reflected in a mirror, as it's shown in figure~\ref{fig:SatAbs_schema}.
The pump saturates the transition and induces transparency, sensed by the probe as
a transmission peak.
This technique enables the possibility of laser frequency locking to absolute references
with higher accuracy.

\begin{figure*}[ht]
  \begin{center}
    \epsfig{file=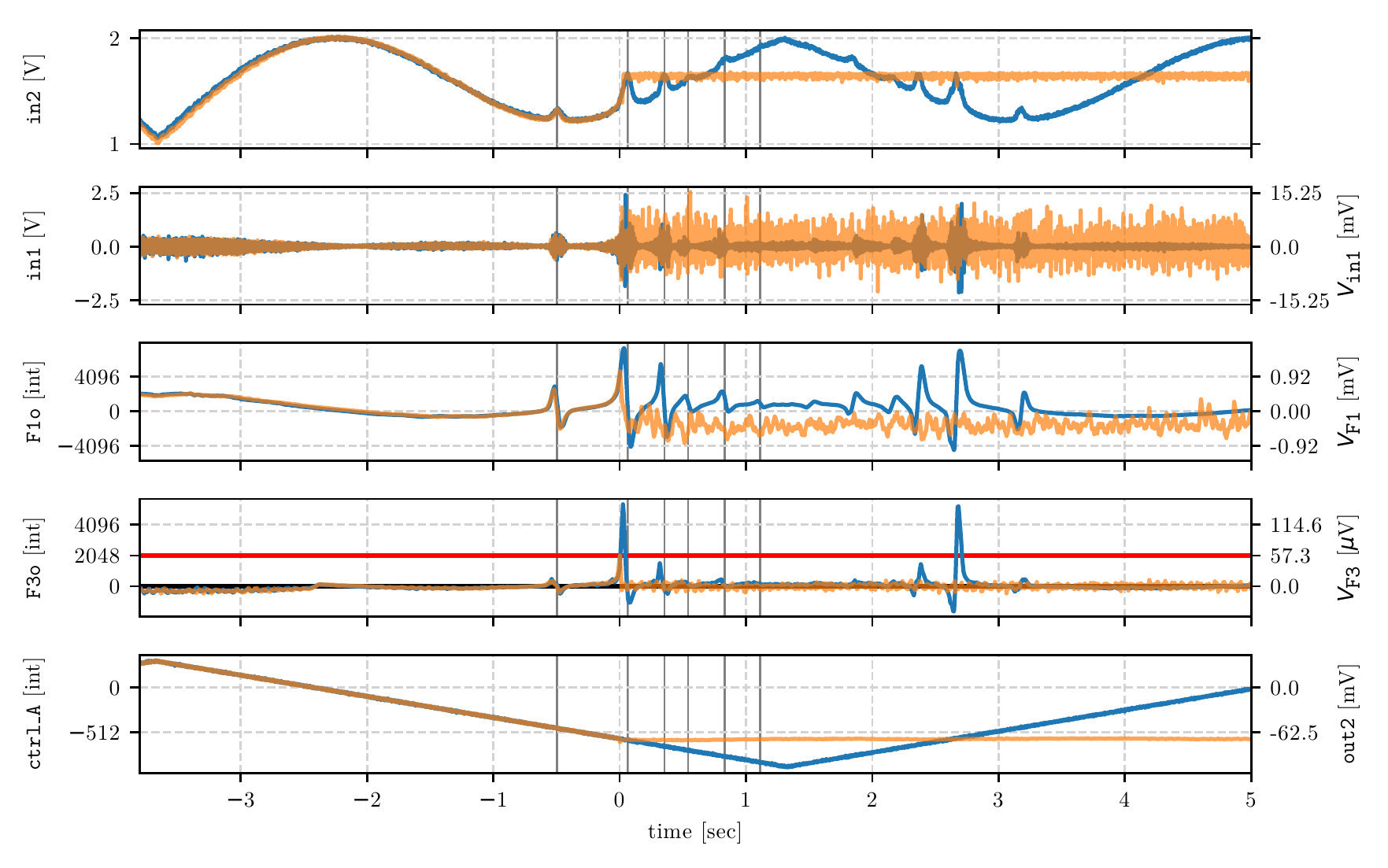, angle=0, width=0.98\textwidth}
  \end{center}
  \vspace{-15pt}
  \caption{\label{fig:Sat_abs_F3}
          Relevant signals for a saturated absorption spectroscopy open-loop scan (\blue)
          and a triggered closed-loop stabilization scheme to an peak maximum (\orange).
          \texttt{in2} is the transmitted signal sensed by the photodiode.
          \texttt{in1} is \texttt{in2} DC-decoupled and amplified (see figure~\ref{fig:abs-exp}), and is used for lock-in demodulation.
          \texttt{F1o} and \texttt{F3o} is the demodulated signals at 1f and 3f respectively.
          \texttt{ctrl\_A} is the control signal, running a Ramp scan.
          The demodulated signals show the first and third derivative-like behavior, with zero-crossing points on \texttt{in2}  peaks maximum.
          \texttt{F3o} was used for \texttt{error} signal and the \red line shows the level threshold for lock-start trigger.
          The left $V_\texttt{signal}$ axis represent the voltage scale of this signals in the photodiode output.
          }
\end{figure*}

We used an ECDL at 780~nm wavelength, suitable for Rubidium $\text{D}_2$ line
($5^2 \text{S}_{1/2} \longrightarrow 5^2 \text{P}_{3/2} $) spectroscopy.

In this kind of lasers, the diode current and the position of the diffraction grating are
controlled to tune the laser. The position of the grating is controlled with a piezoelectric (PZT)
element. If both parameters are controlled simultaneously mode-hop-free tuning range of several
GHz can be achieved\cite{OL_mode_hop_free_Fuhrer}.
Figure~\ref{fig:Sat_abs_F3} shows an example laser frequency scan (\blue line) around the transitions
${F}_\text{g}=3 \longrightarrow {F}_\text{e}=2,3,4 $
for ${}^{85}\text{Rb}$ isotope. In the figure, the optical frequency increases with \texttt{ctrl\_A}.
The sub-Doppler hyperfine transitions peaks and the cross-over\cite{SatAbs_Bhattacharyya2004} peaks
were marked with vertical gray lines.

For the stabilization of the emission frequency a lock-in scheme was
implemented, to get a \texttt{error} signal suitable for peak maximum locking.
and demodulated from the signal sensed by the photodiode.
An harmonic modulation \texttt{cos\_ref} was introduced through current driver.

Two fast inputs and two fast outputs were used in this scheme (figure~\ref{fig:SatAbs_schema}).
The \texttt{in2} port was used for direct photodiode signal digitalization, to measure
the transmitted laser intensity.
The \texttt{in1} port was used to measure AC components of the  photodiode signal,
using an analog high pass filter and amplifier, what allows to improve the sensitivity
of lock-in demodulation.
The acquisition was made using the Oscilloscope instrument, registering the transmitted
intensity and the demodulated signal.
The \texttt{ctrl\_A} signal on \texttt{out1} port was used to control the laser frequency.
The Ramp instrument was used for frequency scanning and PID for frequency stabilization, trough
feedback loop.
The current and PZT sweep amplitudes were tunned through drivers hardware.
The \texttt{out2} port was used to produce the modulation of the laser current.

The Harmonic Lock-in was configured to demodulate the \texttt{in1} signal. Demodulated signals
\texttt{Xo}, \texttt{Yo}, \texttt{F1o}, \texttt{F2o} and \texttt{F3o} were available
for visualization and acquisition.
The first three of them are proportional to the first derivative of the
transmitted signal \texttt{in2} (at first order of modulation depth~\cite{Demtroder2008})
and the last two are proportional to the second and third derivative, respectively.
The odd derivatives expose \texttt{in2} minimum and maximum as zero crossing points.
This characteristic makes them suitable as an \texttt{error} signal for
min/max peak locking.

In figure~\ref{fig:Sat_abs_F3} the \texttt{F1o} and \texttt{F3o} signals are shown.
The \texttt{F3o} is not sensible to first derivatives offsets, like the linear power
increment of the current sweep, and less sensible to peak base line contributions that
shift the minimum / maximum positions of the transmitted signal respect to ideal ones,
which is why it was selected as \texttt{error} signal for the PID input. The filter was configured
with a proportional component with constant $k_p=1.56\,\cdot 10^{-2}$, for fast corrections, and an
integral component with constant $k_i=5.59\,\text{s}^{-1}$ . The Lock-Control instrument was configured to
trigger the loop-close event whenever \texttt{F3o} gets over 2048 int (red line), with the ``level trigger''.
An example of lock start is presented in figure~\ref{fig:Sat_abs_F3}, with \orange line superimposed
to the normal scan signal.
In this example, an stability of $226(13)\,\text{kHz}$ of the optical frequency was achieved after lock,
on a 10 minutes measurement. The frequency deviation was estimated from \texttt{F3o} standard deviation
and the knowledge of the \texttt{F3o} slope on the locked peak position\cite{Luda_OPA_2018}. The \texttt{CTRL\_A} was used
to measure the corrections made to the system, as an estimation of the frequency deviation the laser
would have had if it had worked in open-loop configuration: $54(4)\,\text{MHz}$ RMS over 10 minutes.

\subsection{Pound-Drever-Hall technique:\\* Square Lock-in} 
\label{sec:PDH}

In this experience an ECDL (centered at 854~nm wavelength)  was stabilized to a reflection dip of a
high finesse Fabry-Pérot (FP) cavity using the Pound-Drever-Hall (PDH) technique\cite{PDH_Drever1983}.
This stabilization scheme uses lock-in demodulation of the laser intensity reflected from the
interferometer to produce the \texttt{error} signal.
The modulation frequency must be greater than the cavity's bandwith,
which are both typically chosen to be in the few MHz range.
The laser is modulated producing sidebands that lay out of the transmission peak bandwidth of
the interferometer, so their reflection produces a beating that can be measured by a fast
photodiode. This technique is often used for laser stabilization in AMO Labs\cite{PDH_Drever1983}.

\begin{figure}[hb]
  \begin{center}
    \epsfig{file=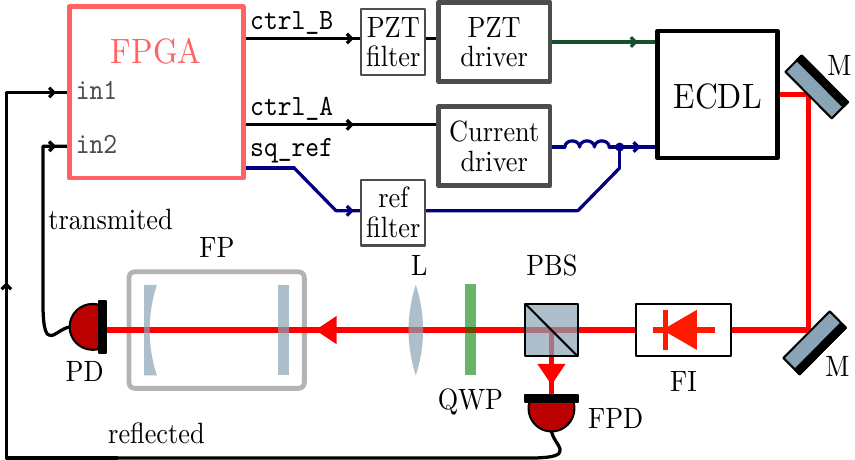, angle=0, width=\columnwidth}
  \end{center}
  \vspace{-15pt}
  \caption{\label{fig:PDH_scheme} Scheme of the Pound-Drever-Hall stabilization technique
            implemented for this work. Includes two control signals for the ECDL and a modulation
            signal (\texttt{sq\_ref}) incorporated directly to the current line with a bias-T configuration.
            The transmitted and reflected signals are collected by photodiodes. The QWP and PBS are combined
            for polarization rotation and reflected beam isolation.
            M: mirror; QWP: quater wave plate; PBS: polarizing beamsplitter;
            FI: Faraday isolator; PD: photodiode; FPD: fast photodiode; L: lens;
            FP: Fabry-Pérot interferometer.}
\end{figure}

\begin{figure}[ht]
  \begin{center}
    \epsfig{file=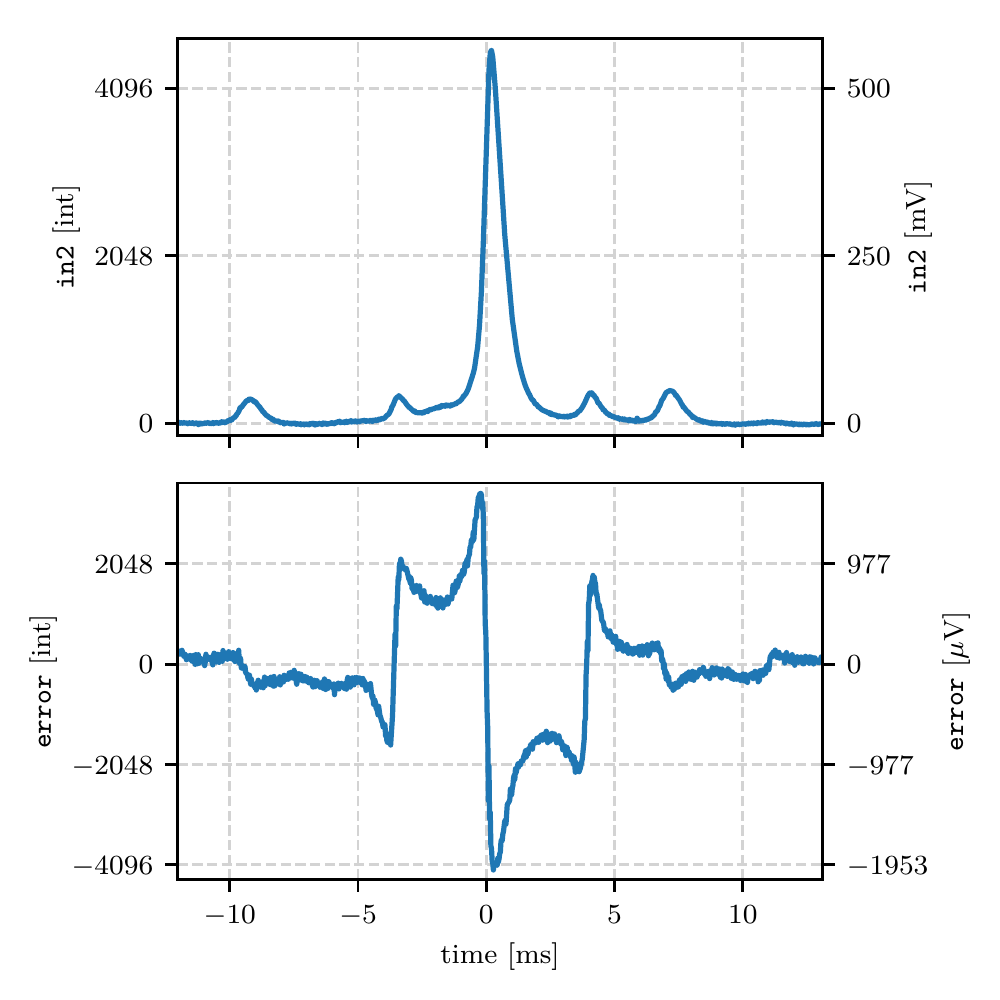, angle=0, width=\columnwidth}
  \end{center}
  \vspace{-15pt}
  \caption{\label{fig:PDH_signal}
            Transmitted beam intensity (top) and the lock-in demodulation of
            reflected intensity (bottom) for a Ramp scan of the Pound-Drever-Hall
            scheme. \texttt{error} signal has not a derivative-like behavior, but produces a good zero-crossing point suitable
            for a PID stabilization control scheme.
            }
\end{figure}

The experimental setup is depicted in figure~\ref{fig:PDH_scheme}.
Two control signals were used for laser control: \texttt{ctrl\_A} for current driver
and \texttt{ctrl\_B} for PZT driver. The radio frequency modulation was incorporated
directly into the laser diode using a bias-T configuration and passive electronic components.
The reflected signal was measured by a fast photodiode with DC-decoupled output and digitalized,
using port \texttt{in1}, for lock-in demodulation. Another photodiode was used to measure
the transmitted signal through \texttt{in2} port for system state reference.
An example measurement for demodulated signal \texttt{error} and \texttt{in2} input
is shown in  figure~\ref{fig:PDH_signal}.

Two configurations of hardware ports were used in the implementation.
Both of them dedicated one of the
14~bits fast outputs (\texttt{out1})  to the current driver. In the first configuration, the other
14~bits fast output (\texttt{out2}) was used for \texttt{sq\_ref} modulation signal and one of the
12~bits slow outputs of the extension bus (\texttt{slow\_out1}) was used for PZT driver.
The slow output update speed is enough for a PZT driver, whose bandwidth limit is in the order of tens
of kHz, but with the downside that the output signal has a high frequency ripple, that comes from
the original PWM signal. To reduce side-effects of the ripple
we included a second order passive low-pass-filter before the PZT driver.
The second configuration option was to use the \texttt{out2} port for the \texttt{ctrl\_B} signal of PZT,
without any filter. In this case, the \texttt{sq\_ref} signal was supplied trough one of the
extension digital pins ($3.3\,\text{V}$ at 125~MSa/sec) as a square function.
A DC-decoupling capacitor and a second order passive low-pass-filter were used to build a band-pass filter,
for signal conditioning: suppression of high frequency components, zero-centered signal and power attenuation.

The \texttt{sq\_ref} modulation signal was configured on $31.25\,\text{MHz}$, the higher available frequency for Square Lock-in.
A second order filter with a frequency cut of $f_c=38.856\,\text{kHz}$ was used for demodulation,
that produces a time lag of $\Delta t \sim 8.2\,\mu\text{s}$ over the \texttt{error} signal.
This imposes an upper bound to the feedback gain and/or to the feedback bandwidth\cite{RMP_Bechhoefer_2005},
but with no fundamental limits to achieve stabilization up to $\frac{1}{2 \Delta t} \sim 61\,\text{kHz}$,
resilient to acoustic and mechanical perturbations.

The Ramp instrument was used for PDH spectrum acquisition, configuring the triangle functions \texttt{ramp\_A}
and \texttt{ramp\_B} with a proportional relation \texttt{ramp\_B\_factor} tuned for optimal mode-hop-free scanning.
The proportional constant was selected considering the hardware configuration option selected.
The feedback scheme was built using one error signal and two PIDs, one for each \texttt{ctrl} signal.
The PID for current setup used proportional and integral terms, while PZT PID only used integral term, avoiding
undesired fast corrections over the piezoelectric line. Sample transmitted and \texttt{error} signal
are presented in figure~\ref{fig:PDH_signal} for one FP transmission peak, showing the characteristic
PDH pattern\cite{PDH_Drever1983}.

Next, we will present two advanced usage cases of the toolkit for continuous data acquisition and
re-lock procedures.

\subsubsection{Allan devition: in-device measurements} 
\label{subsec:allan}
A measurement of the Allan deviation was made to make a detailed analysis of the stabilization performance.
The Allan deviation $\sigma_y(\tau)$ \cite{IEEE_Allan_Standard_Chairman1999} provides a detailed
description of the stability of a system in several orders of magnitude of time. The value $\sigma_y(\tau)$ is
the standard deviation of the differences of successive mean values of $y$, presented in equation~\ref{eq:allan-dev}, where
$y$ is the fractional frequency of an oscillator under study.

\begin{equation}\label{eq:allan-dev}
  \sigma_y(\tau) = \sqrt{ \frac{1}{2} \langle(\bar{y}_{n+1} - \bar{y}_n)^2\rangle  }
\end{equation}

The measurement of $\sigma_y(\tau)$ requires the continuous acquisition of data channels at high sample rate
for long time range. This kind of acquisition cannot be done by the Oscilloscope instrument, limited
on total time range, nor by a remote acquisition procedure, with sample rate limited because of
high communication latency.

\begin{figure}[ht]
    \begin{center}
      \epsfig{file=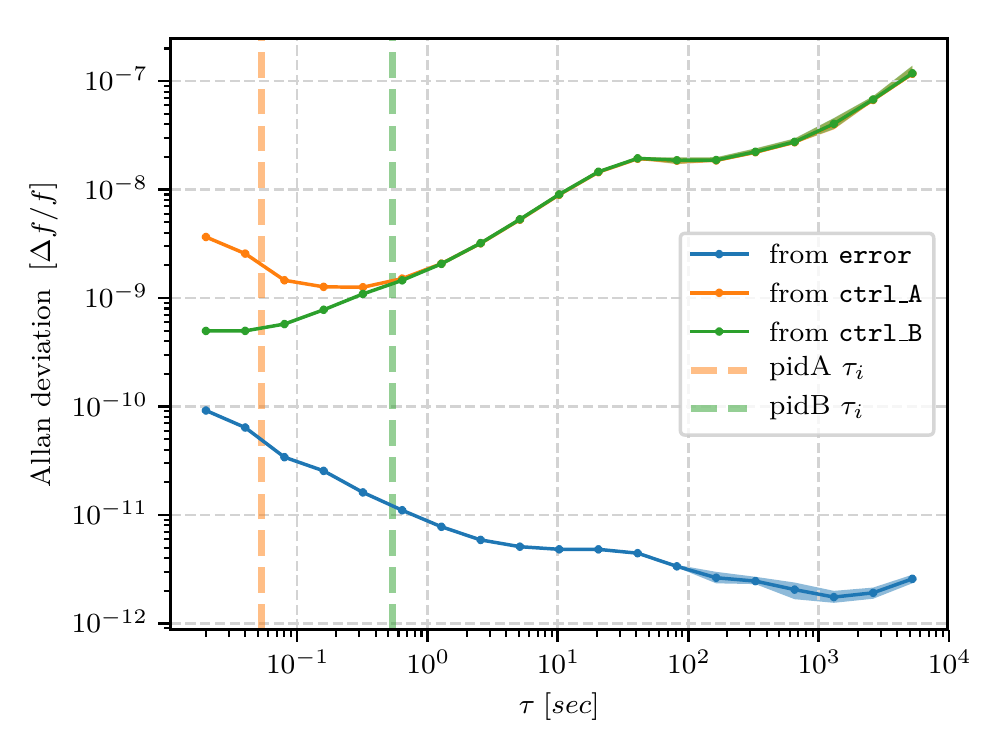, angle=0, width=\columnwidth}
    \end{center}
  \vspace{-15pt}
  \caption{\label{fig:PDH_allan_dev}
            Allan deviation~\eqref{eq:allan-dev} of the fractional frequency of the laser derived from
            \texttt{error} signal (\blue), and from control signals (\orange
            and \green). The first one characterize the performance of the closed-loop
            stabilization system on  different time ranges. The other two are plotted for qualitative comparison:
            the control signals Allan deviations for time ranges $\tau >> \tau_i$ represents the corrections made to the
            system to keep it locked, and can be used as a rough estimation of the open-loop behavior the system would have had
            without stabilization. Dashed lines mark the time constans $\tau_i$ for each PID integrator. At 10~seconds the
            stabilization scheme achieves an improvement of 3 orders of magnitude.
            }
\end{figure}

To make this acquisition we implemented an in-device program, running from a Python script in the
Red Pitaya operative system. The sample script is published in the on-line documentation\cite{ml_github_doc}.
The program consists in a large loop that
reads values directly from RAM memory mapped registers that corresponds to the signals that should be saved and prints them
on the standard output. This simple approach  allows one to redirect the output for local storage or network streaming and
remote storage, using the operative system tools. The RAM memory access and the identification of memory addresses
are simplified by the usage of the local API\@. A set of FPGA registers can be frozen for each read procedure, so all the
acquired values keep coherence (in the sense that they correspond to the same clock time bin). A 64 bit
counter running in the FPGA layer is used to register the accurate internal clock time value of each acquisition.

This implementation allowed to take large measurements of \texttt{error}, \texttt{ctrl\_A} and \texttt{ctrl\_B}
signals with a sample rate of at least $50\,\text{Sa/s}$ along several hours, which were stored in binary files of hundreds of
Mbytes in a remote computer. With this information, the Allan deviation of the fractional frequency of the stabilized laser
(calculated from \texttt{error} signal) was measured, shown in figure~\ref{fig:PDH_allan_dev}. Also, the
\texttt{ctrl\_A} and \texttt{ctrl\_B} signal allow to estimate the open-loop behavior the laser would have had,
by taking into account the corrected deviations during the stabilization time. The fractional frequency Allan
deviation derived from this signals were also plotted. The vertical dashed lines mark the PIDs integrators time
constants asociated with each \texttt{ctrl} signal, as a reference. For $\tau_i$ larger than these references, the
curves derived from \texttt{ctrl} signals can be interpreted as frequency corrections. The increment on
\texttt{ctrl\_A} derived values on short times is related with the proportional term of the PID used for current control,
and tends to reflect the behavior of \texttt{error} signal.
An improvement of three orders of magnitude in the stability at 1~second time range can be seen from the plot.

\begin{figure}[ht]
    \begin{center}
      \epsfig{file=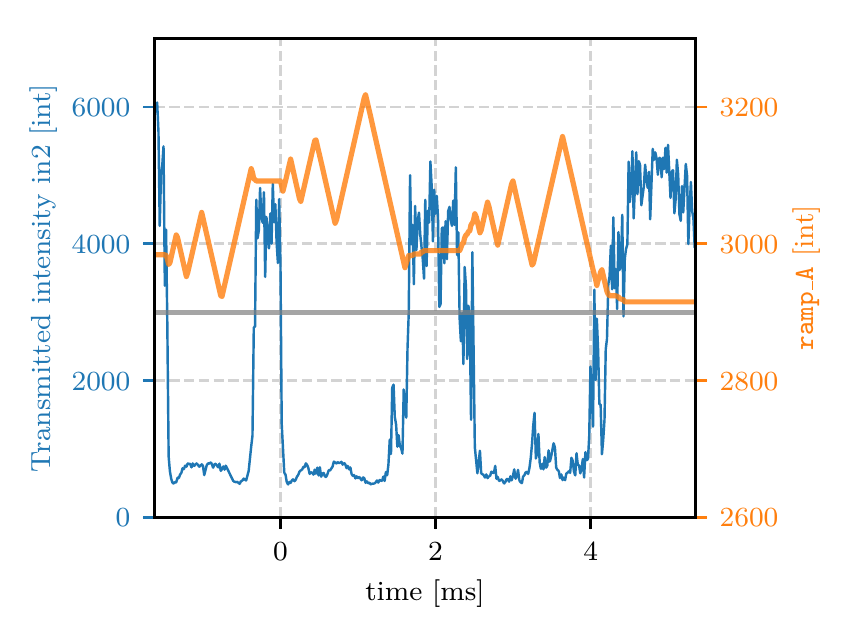, angle=0, width=\columnwidth}
    \end{center}
  \vspace{-15pt}
  \caption{\label{fig:PDH_relock}
            Relocking feature of Control-lock instrument. In this example the relock system was configured to start
            when \texttt{in2} signal falls below $3000\,\text{int}\,\cong\,366\,\text{mV}$ (\gray~line). The triangular sweep
            increase the scan amplitud on each semi-period until  it reachs the lock condition again. This feature is
            described in subsection~\ref{subsec:lock-control_intrument}.
            }
\end{figure}

\subsubsection{Re-lock system} 
\label{subsec:relock}

The Lock-control instrument includes the feature to identify locking events and actuate on them, already described in
subsection~\ref{subsec:lock-control_intrument}. The PDH example provides a case study to test it. With the system locked
to a transmission peak, the re-lock tool was configured to trigger when $\texttt{error}>1000\,\text{int}$ or when
transmitted signal $\texttt{error}<3000\,\text{int}\cong 366\,\text{mV}$. The ``Out of lock'' external trigger allowed to capture the
re-lock system response under an  stabilization induced fail, by hard hitting the optic table, what
is shown in figure~\ref{fig:PDH_relock}. Three cycles of re-locking can be seen, while the
mechanical vibrations are still affecting the system.

\section{Conclusion} 

We presented an embedded toolkit for digital processing, acquisition and feedback control
through MIMO control design. The combination of FPGA fast deterministic-timing for signal
processing and microprocessor with operative system for overall monitoring and control provides
a balance between programmable electronic precision and algorithm versatility.
This allowed the implementation of several usage strategies going from simple out-of-the-shelve
gross-control to in-device programed fine-control, as shown the experimental examples.

The in-device operative system provides portability and multi-platform GUI access through web browser.
The PIDs, designed to work on several orders of magnitude, enable the usage on several control applications, even beyond the ones belonging to AMO labs presented in this work.
The design of two lock-in instruments enabled usage for precision measurements and for large
working frequency range, including the possibility of the implementation of the complete
PDH modulation and lock-in demodulation at $31.25\,\text{MHz}$ in one device, something that
had not been reported so far.

The selection of an economical commercial board\cite{redpitaya_STEMLab} may ease the acquisition
and fast implementation of this toolkit by a third party in new experiments. The toolkit
FPGA design and software are in public domain, and even the board operative system and applications framework
are open-source, what encourages others to use, modify and add new features or bugfix. Also, the compact design
and remote programmable feature make the toolkit useful for mass implementation on experiments with several
control systems with centralized monitoring and operation.

\section*{Acknowledgments} 

We thank Ferdinand Schmidt-Kaler for opening the doors of the Cold ions and
quantum information laboratory (Institut für Physik, Mainz) where the first
laser stabilization tests with FPGA where conducted.
The financial support for this research was provided by the
Ministry of Defense of Argentina (PIDDEF~23/14)
and the National Agency of Promotion of Science and Technology
(PICT~2014-3711.11).




\bibliography{referencias}

\end{document}